\documentclass[conference]{IEEEtran}
\usepackage[pdftex]{graphicx}
\DeclareGraphicsExtensions{.pdf,.jpeg,.png}
\usepackage[utf8]{inputenc}
\usepackage{subcaption} 
\usepackage{hyperref}
\usepackage{mathtools}
\usepackage{smartdiagram}
\usepackage{empheq}
\usepackage[most]{tcolorbox}
\usepackage[normalem]{ulem}
\usepackage{amssymb}
\usepackage{scrextend}
\usepackage{xcolor}
\newtheorem{remark}{Remark}

\begin{document}
\title{\LARGE \bf Secure, Anonymity-Preserving and Lightweight Mutual Authentication and Key Agreement Protocol for Home Automation IoT Networks} 
\author{\Large Akash Gupta, Gaurav S. Kasbekar}
\maketitle
\thispagestyle{plain}
\pagestyle{plain}


{\renewcommand{\thefootnote}{} \footnotetext{A. Gupta is with the Defence Electronics Application Laboratory (DEAL), Defence Research \& Development Organization (DRDO), Dehradun, India.  G. S. Kasbekar is with the Department of Electrical Engineering, Indian Institute of Technology (IIT) Bombay, Mumbai, India. Their email addresses are akash.iitb19@gmail.com and gskasbekar@ee.iitb.ac.in, respectively. A. Gupta worked on this research while he was with IIT Bombay.}}

\begin{abstract}
Home automation Internet of Things (IoT) systems have recently become a target for several types of attacks. In this paper, we present an
authentication and key agreement protocol for a home automation network based on the ZigBee standard, which connects together a central controller and several
end devices. Our scheme performs mutual authentication between
end devices and the controller, which is followed by device-to-device
communication. The scheme achieves confidentiality, message integrity, anonymity, unlinkability, forward and backward secrecy, and availability. Our
scheme uses only simple hash and XOR computations and symmetric
key encryption, and hence is resource-efficient. We show using a detailed security analysis and numerical results that our proposed scheme provides better security and anonymity, and is more efficient in terms of computation time, communication cost, and storage cost than schemes proposed in prior works. 
\end{abstract}

\IEEEoverridecommandlockouts
\IEEEpeerreviewmaketitle

\section{Introduction}
Recently, there has been a proliferation of Internet of Things (IoT) systems, which include a large number of low-cost devices equipped with different types of sensors, transceivers and microcontrollers, for enabling a variety of applications~\cite{al2015internet}. One such example is \emph{home automation IoT systems}, which provide better control over lighting, windows, doors, thermostats, and other appliances in a household~\cite{al2015internet}. 
In a home automation IoT system, various low-cost IoT devices frequently need to communicate with each other. However, recently, there have been several instances of attacks on IoT systems by intruders, e.g., information theft,  Denial-of-Service (DoS) attacks, etc.~\cite{hofer2020towards}. Hence, the communication among IoT devices in a home automation system
needs to be secure. Also, since the  battery availability, storage, computational, and communication capabilities of IoT devices are limited, the security mechanisms used must have a low complexity, and conventional security protocols designed for traditional Internet-connected devices such as Desktop computers and laptops cannot be used~\cite{al2015internet}.

Security has been a focal point of research in IoT in recent years and many protocols have been proposed for the same~\cite{granjal2015security}. However,  these schemes primarily use \emph{static identities} for devices. In a	home automation environment, the roles of most of  the appliances are fixed. For example, a coffee machine can only prepare coffee, and a door opening system can only open or close the door. If the static identities of the devices involved (e.g., devices attached to coffee machine, door opening system) are disclosed, it is easy for an intruder to guess the action intended in a small home automation network without even decrypting the payload. 
Also, recently, there have been attacks on baby heart monitors, which are end devices in the IoT health care infrastructure~\cite{zhou2018effect}, in which intruders impersonated legitimate devices; such attacks can have serious consequences. Thus, it is of paramount importance to design an \emph{authentication and key agreement protocol}, which is efficient and suitable for a constrained environment, secure against various types of advanced attacks, and uses \emph{dynamic identities}. 

In this paper, we design such a protocol for a home automation network based on ZigBee, which is a popular wireless networking standard~\cite{callaway2002home}.
We consider a two-tier ZigBee based home automation scenario with a central controller and several end devices. In our protocol, the real identities of the controller and devices are kept secret; a counter,  which is known only to the parties exchanging a message and increments after every message exchanged,  is used to derive dynamic identities for every message exchanged. Hence, even a malicious insider cannot trace the communication. The payload of every message is encrypted using a symmetric key and message integrity is achieved using a secure one-way collision-resistant keyed hash function~\cite{pfitzmann2001anonymity}. Our scheme performs mutual authentication between end devices and the controller followed by device-to-device (D2D) communication. The scheme also achieves anonymity,  unlinkability, forward and backward secrecy, and availability~\cite{pfitzmann2001anonymity},~\cite{grammatikis2019securing}. At the same time, the scheme uses only simple hash and exclusive-OR (XOR) computations and symmetric key encryption, and hence is resource-efficient.

The rest of the paper is organized as follows. We provide a review of related literature in Section~\ref{RW}. In Section~\ref{NM}, we describe the network model, problem formulation, relevant background, and attacker model. We describe the proposed scheme in Section~\ref{PS}. In Section~\ref{SA}, we provide a security analysis of the proposed scheme and qualitatively compare it with schemes proposed in prior works. We provide numerical results in Section~\ref{CA} and conclude the paper in Section~\ref{CONC}.

\section{Related Work}
\label{RW}

In~\cite{RF:Perrig:TESLA}, the TESLA protocol was proposed for solving the broadcast authentication problem. However, it does not provide the crucial properties of anonymity and unlinkability, which our proposed scheme provides. 

Several schemes have been proposed for secure device-to-device communication in IoT systems. In~\cite{yeh2011secured}, the authors proposed an elliptic curve cryptography (ECC) based user authentication protocol for Wireless Sensor Networks (WSN). Later, in~\cite{turkanovic2014novel}, the authors showed that the ECC based system proposed in~\cite{yeh2011secured} requires a large amount of resources and proposed a lightweight and energy-efficient scheme that relies only on simple hash and XOR computations for mutual authentication between two parties. Their scheme was the first to consider a user contacting a sensor node directly, while prior schemes involved a user contacting the gateway node for mutual authentication and then getting a session key for communication with a sensor node. In~\cite{amin2016secure},  several security weaknesses of the scheme presented in~\cite{turkanovic2014novel} were demonstrated. The authors in~\cite{amin2016secure} pointed out that public key cryptosystems such as RSA, ECC, El-Gamal, etc., have a high computational cost and hence are not suitable for energy-constrained WSNs. Our proposed scheme does not use public key cryptography. In~\cite{risalat2017advanced}, a hash-based mutual authentication protocol for radio frequency identification (RFID) systems was presented. 
In~\cite{wazid2019design}, an authentication and key agreement scheme was proposed for fog computing services, which uses lightweight operations such as a one-way cryptographic hash function  and bitwise XOR operations. The schemes proposed in~\cite{yeh2011secured},~\cite{turkanovic2014novel},~\cite{amin2016secure},~\cite{risalat2017advanced},~\cite{wazid2019design} are designed for WSNs, RFID systems or fog computing services; in contrast, in this paper, we propose an authentication and key agreement protocol for home automation systems.

In~\cite{bamasag2015towards}, a novel continuous authentication protocol for the IoT based on hash functions, message authentication codes, and a secret sharing scheme was presented. Similarly, in~\cite{lee2020three}, the authors proposed a three-factor user authentication scheme for the IoT. However, both the above protocols need clock synchronization, which requires overhead to achieve. Our proposed scheme does not require any clock synchronization. In~\cite{avoine2019iot}, the authors presented a three-party authenticated key exchange protocol solely based on symmetric-key functions. They consider a LoRaWAN-like architecture with a trusted third-party server, which increases the infrastructure overhead. Our proposed scheme does not require any third-party server. Moreover, the schemes proposed in~\cite{bamasag2015towards,avoine2019iot} do not provide the properties of anonymity and unlinkability, which our proposed scheme provides. 

In~\cite{ashibani2017context}, a context-aware authentication framework for smart homes was introduced, which utilizes contextual information such as the user's location, profile, calendar, request time, and access behavior patterns to enable access to home devices. The introduced framework protects smart devices against unauthorized access by anonymous and known users, whether local or remote, by routing all communication to the devices through a secure gateway. However, the scheme proposed in~\cite{ashibani2017context} does not provide authentication in device-to-device communication and also does not satisfy the properties of anonymity and unlinkability~\cite{alshahrani2019anonymous}.

In~\cite{alshahrani2019anonymous}, the authors proposed a scheme based on an incremental counter for smart home networks that have a two-tier architecture with a central controller and end devices. Their scheme relies upon an accelerometer reading for key and counter generation in each instance and random nonce techniques. This implementation avoids the need for clock synchronization and prevents replay attacks. The scheme allows IoT devices to mutually authenticate and communicate anonymously in an unlinkable manner. In~\cite{lohachab2019ecc}, the authors proposed an authorization and authentication scheme based on ECC and access control mechanisms. However, the scheme proposed in~\cite{lohachab2019ecc} cannot handle device dynamics such as devices joining or leaving the network~\cite{pampapathi2021data}; in contrast, our proposed scheme can handle device dynamics. Also, unlike the scheme proposed in~\cite{lohachab2019ecc}, our proposed scheme does not use public key cryptography and hence is faster and more efficient. In Sections~\ref{SSC:comparison:prior:work} and~\ref{CA}, we show using detailed qualitative analysis and numerical results that our proposed scheme provides better security and anonymity, and is more efficient in terms of computation time, communication cost, and storage cost than those proposed in~\cite{alshahrani2019anonymous} and~\cite{lohachab2019ecc}.


\section{Network Model, Problem Formulation And Background}	
\label{NM}

\begin{figure}[!h] 
\centering
\includegraphics[scale=0.4]{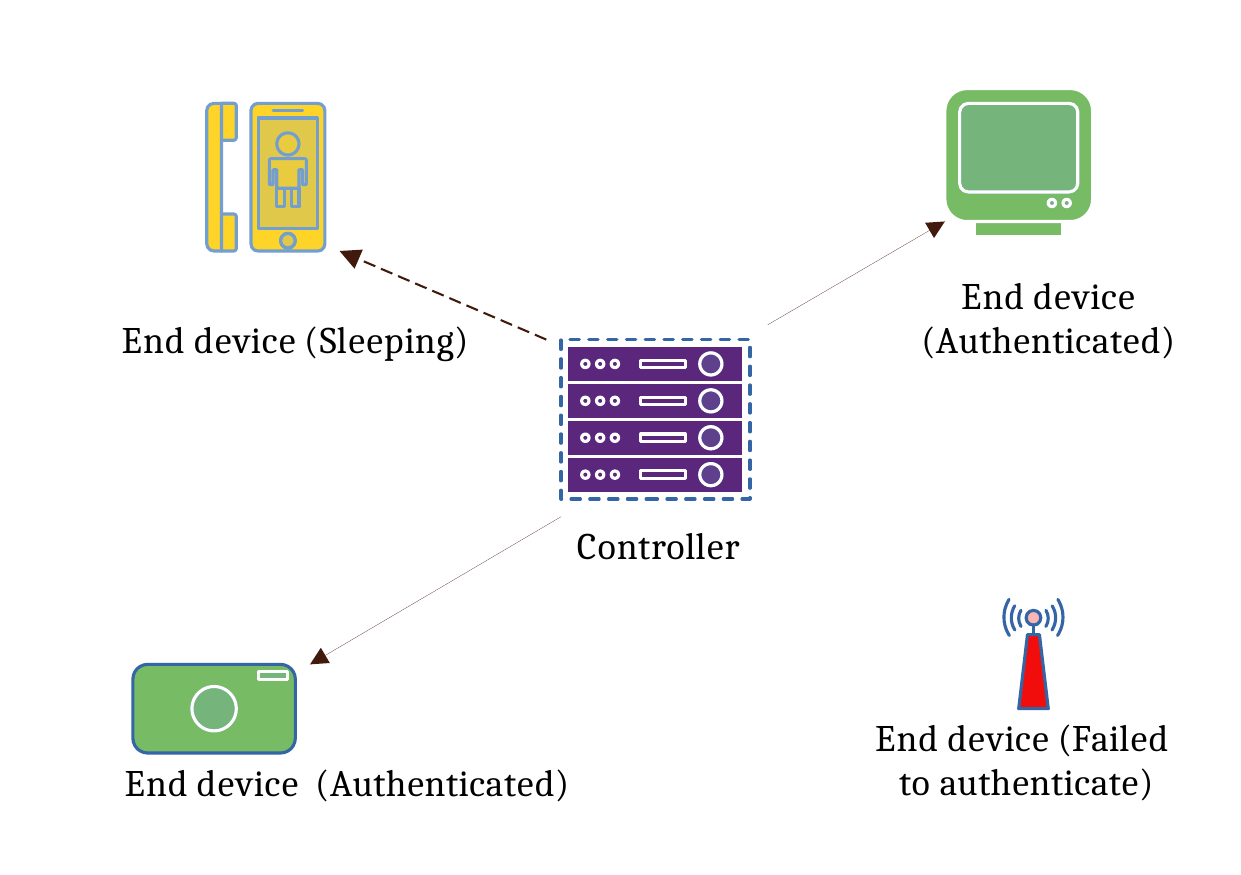}
\caption{The figure illustrates the network model, which includes a  controller and multiple end devices.}
\label{fig1}
\end{figure}


\subsection{Network Model}
ZigBee is a popular wireless networking standard based on IEEE 802.15.4~\cite{callaway2002home}, which is suitable for a constrained environment~\cite{wang2019security}. We consider a two-tier ZigBee based home automation scenario with a central controller and several end devices as illustrated in Fig.~\ref{fig1}. The central controller, denoted as $C$, has high computational, storage, and battery capabilities, while the end devices are constrained in terms of these capabilities. $C$ is in a role similar to that of a \textit{ZigBee Coordinator}, and the end devices are in a role similar to that of \textit{ZigBee End Devices}~\cite{fan2017security}. 

In our setup, a pair of devices can communicate with each other only if the controller allows them to do so. Every communication instance is controlled and directed by the controller. When a new device is added to the network, the controller provides its real identity to the rest of the devices and updates the access control rule. When an old device is removed from the network, the controller  deletes that device from the access control list and notifies the other devices.

Many authentication protocols proposed in prior work use static identities, which are often sent in plaintext form and reveal information to a potential intruder about the identities of entities exchanging messages in the system \cite{turkanovic2014novel}, \cite{amin2016secure}, \cite{risalat2017advanced}, \cite{lohachab2019ecc}. If an intruder captures a message, he/ she can find out as to which two devices are communicating. In a smart home network, each end device has a specific role out of a limited set of roles (e.g., exchanging messages relevant to a door opening system or a coffee machine) and it is easy for an intruder to find out the intended action of a message from the identities of the devices exchanging the message. Prevention of the disclosure of the identities of devices, thus, becomes crucial~\cite{khanji2019zigbee}.

Many authentication schemes use timestamps to prove the freshness of messages~\cite{yeh2011secured},~\cite{turkanovic2014novel},~\cite{amin2016secure},~\cite{risalat2017advanced},~\cite{wazid2019design},~\cite{lohachab2019ecc}. Timestamps are included in every message from the sender, and the receiver discards a message if the time difference between reception and transmission is more than a threshold. But for measuring this time difference accurately, such schemes need precise time synchronization among all devices, which requires additional overhead to achieve. If synchronization fails, an attacker can launch replay attacks, using the fact that the network cannot reliably distinguish between fresh and old packets. Hence, in this paper, we seek to design an authentication protocol that does not use timestamps,  which will eliminate the overhead required to maintain time synchronization among different devices. 


\subsection{Security Design Goals}
\label{SSC:security:design:goals}
Recently, the attack arsenal for IoT systems has been increasing at a significant pace. Various surveys like~\cite{grammatikis2019securing} and~\cite{zhao2013survey} present different threats faced by IoT systems. We aim to design a robust and secure authentication and key agreement scheme that  achieves the following goals:

\begin{enumerate}[\setlength{\IEEElabelindent}{\dimexpr-\labelwidth-\labelsep}
    \setlength{\itemindent}{\dimexpr\labelwidth+\labelsep}
    \setlength{\listparindent}{\parindent}
  ]
\item{The scheme should be able to provide confidentiality. The data exchanged between two devices using the proposed scheme should not be revealed to any intruder~\cite{grammatikis2019securing}.}

\item{The scheme should be able to protect the integrity of the data. An altered message should be detected and discarded~\cite{grammatikis2019securing}.}

\item{The scheme should be able to provide anonymity to the participating devices. A  device should be able to communicate with any other device in the network without revealing its real identity to anyone except the receiver. Devices should use purely dynamic identities to remain completely anonymous~\cite{grammatikis2019securing}.}

\item{Home automation networks may need the addition of a new device (e.g., one with desired features) or removal of a device (e.g., malfunctioning one). The scheme should enable dynamic entity removal (respectively, addition) with forward (respectively, backward) secrecy.  That is, the scheme should prevent newly removed (respectively, added) devices from getting any information from any future message exchanges in the network (respectively, understanding past messages)~\cite{kabra2020efficient}.}

\item{The scheme should be able to maintain \emph{unlinkability}~\cite{pfitzmann2001anonymity}. That is, if three devices, say $N_1$, $N_2$ and $N_3$, are communicating with each other pairwise, then $N_3$ should not be able to find out that any intercepted message was sent by $N_1$ to $N_2$ or vice-versa, although it is communicating with both of them.}
\end{enumerate}

In addition, the proposed scheme should be able to defend against replay attacks, man-in-the-middle attacks, Sybil attacks, DoS attacks, identity theft attacks, timing attacks, side-channel attacks, and impersonation~\cite{butun2019security}. The scheme should also address the issues of session expiry, session key guessing, and parallel sessions~\cite{hofer2020towards}. In this work, we do not consider node capture and physical-layer attacks.
\begin{center}
\begin{table}[h!]
\normalsize
\caption{Abbreviations and notation used in this paper}
\begin{tabular}{ |l |l |}
\hline
\textbf{Notation}  & \textbf{Description}  \\ 
\hline
$SA$  &  System Administrator \\\hline
$C$  & Controller \\\hline
$N_i$  & IoT end device $i$ \\\hline
$ID_{i}$  &  Real identity of device $N_i$ \\\hline
$AUTH\_REQ$ & Authentication Request for a new session \\\hline 
$r_i$  &  Random number generated by $N_i$ \\\hline
$CC_{i}$  & The counter shared between $N_i$ and $C$ \\\hline
$h(\cdot)$  & Hash function \\\hline
$CC_{i,L}$  & Last $L$ bits of $h(CC_i)$\\\hline
$K_{i}$  & Symmetric key with $C$ for device $i$ \\\hline
$HMAC$  & Keyed-hash message authentication code  \\\hline
$DDC_{ij}$   & The counter shared between $N_i$ and $N_j$ \\\hline
$TK_{ij}$  &  Temporary shared key between $N_i$ and $N_j$ \\\hline
$MI^{C}_{i}$  & Masked identity of $N_i$ for Controller \\\hline
$MI^{i}_{C}$  & Masked identity of Controller for $N_i$ \\\hline
$MI^{j}_{i}$  & Masked identity of $N_i$ for $N_j$  \\\hline
$OTP_{i}$  & One-Time Password for $N_i$  \\\hline
$PoB_{i}$  &  Proof of Belonging for device $N_i$ \\\hline
$PoC$  & Proof of Controller  \\\hline
$\parallel$  &  Concatenation \\\hline
$\oplus$  &  Bitwise XOR \\\hline
\end{tabular}
\label{table:t0}
\end{table}
\end{center}
\vspace{-.75cm}


The abbreviations and notation used in this paper are provided in Table~\ref{table:t0}. $ID_i$ is the real identity of device $N_i$, provided by the $SA$. $AUTH\_REQ$ is a fixed bit sequence, which indicates that this message is intended for authentication. $HMAC$ is computed using the present value of the shared counter as the key. $PoB$ and $PoC$ are hash values computed using certain parameters and presented to prove the fresh belongingness to the network.

\subsection{Attacker Model}
Our goal is to design a scheme that is secure against an attacker who has the capabilities and limitations defined in the \emph{Dolev-Yao's (DY) threat model}~\cite{dolev1983security}. According to this model, the scheme is said to be secure if the adversary cannot decrypt a message without the encryption key ($K_{i}$ or $TK_{ij}$), cannot compute a keyed $HMAC$ without the key ($CC_{i}$ or $DDC_{ij}$), and cannot guess an encryption key ($K_{i}$), a nonce ($CC_{i}$), or a random number ($r_i$). 

The adversary  is assumed to be aware of all the public data of the protocol; he/ she is also able to read, store, and block every message in transit. The adversary can create and transmit new messages. He/ she can benefit from all the privileges/ keys of bad agents. The adversary is able to encrypt/ decrypt if the encryption/ decryption key is known. He/ she is able to initiate any number of parallel protocol sessions.

\section{Proposed Scheme}
\label{PS}
In our proposed scheme, a secret pre-shared bit string ($p_i$) is used in combination with a session-specific random number ($r_i$) to calculate a fresh unique counter ($CC_i$) and symmetric key ($K_i$) for every session. The random number $r_i$ is generated at end device $i$, encrypted and sent to the controller in an authentication request. A deterministic  pseudo-random function is used to derive the counter ($CC_i$) and the session key ($K_i$) from a given pair $(p_i, r_i)$. The quantity $p_i$ is specific to the device $N_i$ and is known only to the controller and device $N_i$ itself. This fact prevents an insider, with the knowledge of $ID_i$, from masquerading as $N_i$  as he/ she also needs $p_i$  for authenticating with the controller.  

Our scheme is divided into three phases:  (A) Registration phase, (B) Authentication phase, and (C) D2D Communication phase; these phases are described in Sections~\ref{SSC:registration:phase},~\ref{SSC:authentication:phase} and~\ref{SSC:D2D:communication:phase}, respectively.

\subsection{Registration Phase }
\label{SSC:registration:phase}
During the registration phase, the system administrator ($SA$) configures all devices by storing their real identities ($ID_i$) and bit-strings ($p_i$) in the  controller ($C$). $p_i$ is a randomly generated secret bit-string, which is pre-shared between device $N_i$ and the controller. $p_i$ is used by $C$ to identify different registered IoT devices and differentiate between them during the authentication process. The real identities ($ID_i$) of the IoT devices and the controller's identity ($ID_C$) are kept secret and never transmitted in plaintext form; instead, only dynamic IDs are used during authentication and communication. During registration, end devices are initialized with some random values of $CC_{i}$ and $K_{i}$ by the $SA$ and these values are also shared with the controller to enable the first authentication of the device to the controller. 

\subsection{Authentication Phase}
\label{SSC:authentication:phase}
Suppose end device $N_1$ wants to authenticate itself with the controller $C$. We assume that $N_1$ has been registered with the controller previously and the 
counter value $CC_{1}$, symmetric key  $K_{1}$, and bit-string $p_1$ have been shared between $N_1$ and $C$. If $N_1$ had previously authenticated with the controller, then the previous session's counter $CC_{1}$ and key $K_{1}$ are stored in the databases of $N_1$ and $C$. 

The general format for message exchange in the proposed scheme is $<\mbox{To (Receiver ID), Message, HMAC}>$.

After a message exchange (with $C$ or with other end devices), if any device remains dormant for a certain time duration, then it has to again  initiate mutual authentication with the controller for a new session as explained in step A1 below before it can engage in further communication. Also, $C$ securely sends an update message to all the connected devices, informing them whenever a new device joins or a device leaves the network.  

\begin{figure}[!h] 
\centering
\includegraphics[scale=0.45]{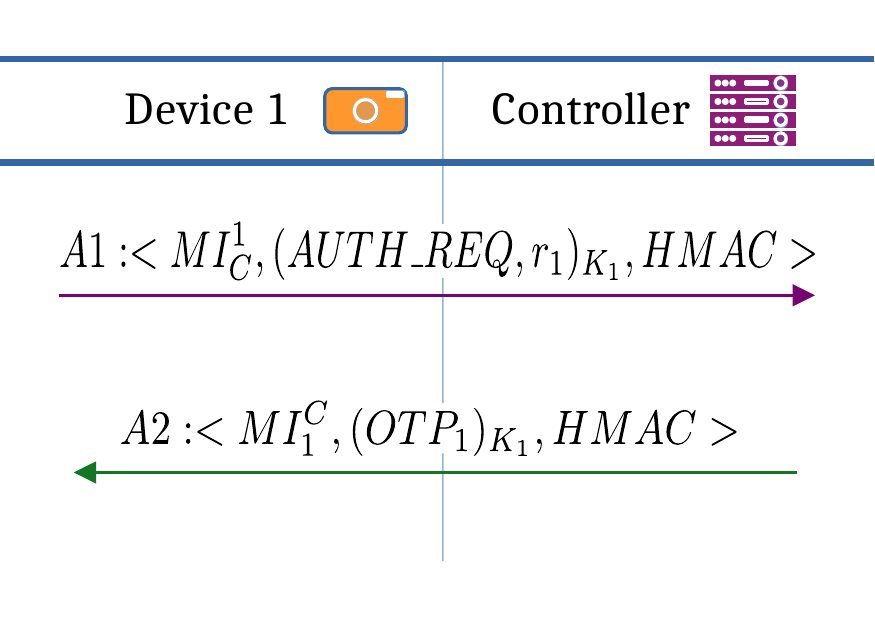}
\caption{Mutual authentication between a device and the controller.}
\label{fig2}
\end{figure}

The authentication phase consists of two steps: A1 and A2, which are shown in Fig.~\ref{fig2} and described below. 

\textbf{Step A1:} The device $N_1$ sends the following message to the controller: 
\begin{center}
$<MI^{1}_C, (AUTH\_REQ , r_1)_{K_{1}}, HMAC>,$ \\
\end{center}
where in the receiver ID field, $MI^{1}_C$ is the masked identity of the controller for an $N_1\leftrightarrow C$ exchange, calculated using the counter value $CC_{1}$ from the previous session (if any) or the counter value $CC_{1}$ shared during registration (if this is the first session), as $h(CC_{1},ID_C)$. 

The controller has its masked identity $MI^{i}_C$ for every device $N_i$ calculated and stored in its database. It accepts a message if the receiver field matches with the calculated masked identity for any device. E.g., when the controller finds a match of the receiver field  $MI^{1}_C$ with the masked identity for device $1$ in its database, it concludes that the message has been sent by device $1$. This makes our scheme efficient as we do not have to include the sender $ID$ in a message. 

In the payload, $AUTH\_REQ$, a fixed string, indicates that this message is an authentication request from an end device to the controller. $p_1$ and $r_1$ are used by the device $N_1$ as inputs to a pseudo-random function to derive the fresh values of $CC_{1}$ and $K_{1}$; these fresh values are updated by $N_1$ in its database. $HMAC$ in this step is $h(CC_{1},[MI^{1}_C \parallel (AUTH\_REQ , r_1)_{K_{1}}])$.

After accepting this message, the controller first calculates $HMAC^* = h(CC_{1},[MI^{1}_C \parallel (AUTH\_REQ,r_1)_{K_{1}}])$ using $CC_{1}$ from its database. If $HMAC^*$ does not match $HMAC$, the message is discarded. The controller uses the previous session's key $K_{1}$ to decrypt the payload. The controller gets the same fresh values of $CC_{1}$ and $K_1$ using the pseudo-random function as $N_1$ gets, assuming that it uses the correct values of $p_1$ and $r_1$. The new values of $CC_{1}$ and $K_1$ are updated in the database. Also, after sending this message, $N_1$ calculates the new $MI^{C}_{1} = h(CC_{1},ID_1)$ using the fresh $CC_1$ and updates it in its database. 

\textbf{Step  A2:} After the controller derives fresh values of $CC_{1}$ and $K_1$, it sends the following message to the device:
\begin{center}
$<MI^C_1, (OTP_1)_{K_1}, HMAC>,$
\end{center}
where $MI^{C}_{1} = h(CC_{1},ID_{1})$ is the masked identity of $N_1$,  calculated using the fresh value of $CC_1$. A one time password, $OTP_{1}$, is generated at the controller and is encrypted using the freshly derived symmetric key $K_{1}$. $OTP_1$ is later used by the device $N_1$ to prove its fresh belonging to the network. In this message, $HMAC$ is $h(CC_{1},[MI^{C}_{1} \parallel (OTP_{1})_{K_{1}}])$, where $CC_1$ is the fresh counter value for this session.

After sending this message, the controller increments the value of $CC_{1}$ by $1$. Next, the controller calculates $MI^{1}_{C} = h(CC_{1},ID_C)$ and $MI^{C}_{1} = h(CC_{1},ID_{1})$ by using the incremented value of $CC_{1}$ and 
updates these in its database. The controller will use these masked identities the next time it communicates with $N_1$.

The end device $N_1$ accepts this message if the receiver field, $MI^{C}_{1}$,  
equals $h(CC_{1},ID_{1})$, where $CC_{1}$ is the value in $N_1$'s database. $N_1$ calculates $HMAC^* = h(CC_{1}, [MI^{C}_{1}, (OTP_{1})_{K_{1}}])$. If $HMAC^*$ matches with $HMAC$, then $N_1$ considers the message to be unaltered and uses $K_{1}$ to decrypt $OTP_1$. $OTP_{1}$ is later used by $N_1$, at the time of D2D communication,  to prove its fresh belongingness to the controller. 

After storing $OTP_1$ in its database, the device $N_1$ increments $CC_{1}$ by  $1$, calculates $MI^{C}_{1} = h(CC_{1},ID_{1})$ and $MI^{1}_{C} = h(CC_{1},ID_{C})$ using the incremented value of $CC_{1}$, and updates them in its database. $N_1$ will use these masked identities the next time it communicates with the controller.

With the above two steps, A1 and A2, the mutual authentication between the controller and device $N_1$ is complete. 


\subsection{D2D Communication}
\label{SSC:D2D:communication:phase}
\begin{figure}[!h]
\centering
\includegraphics[scale=0.4]{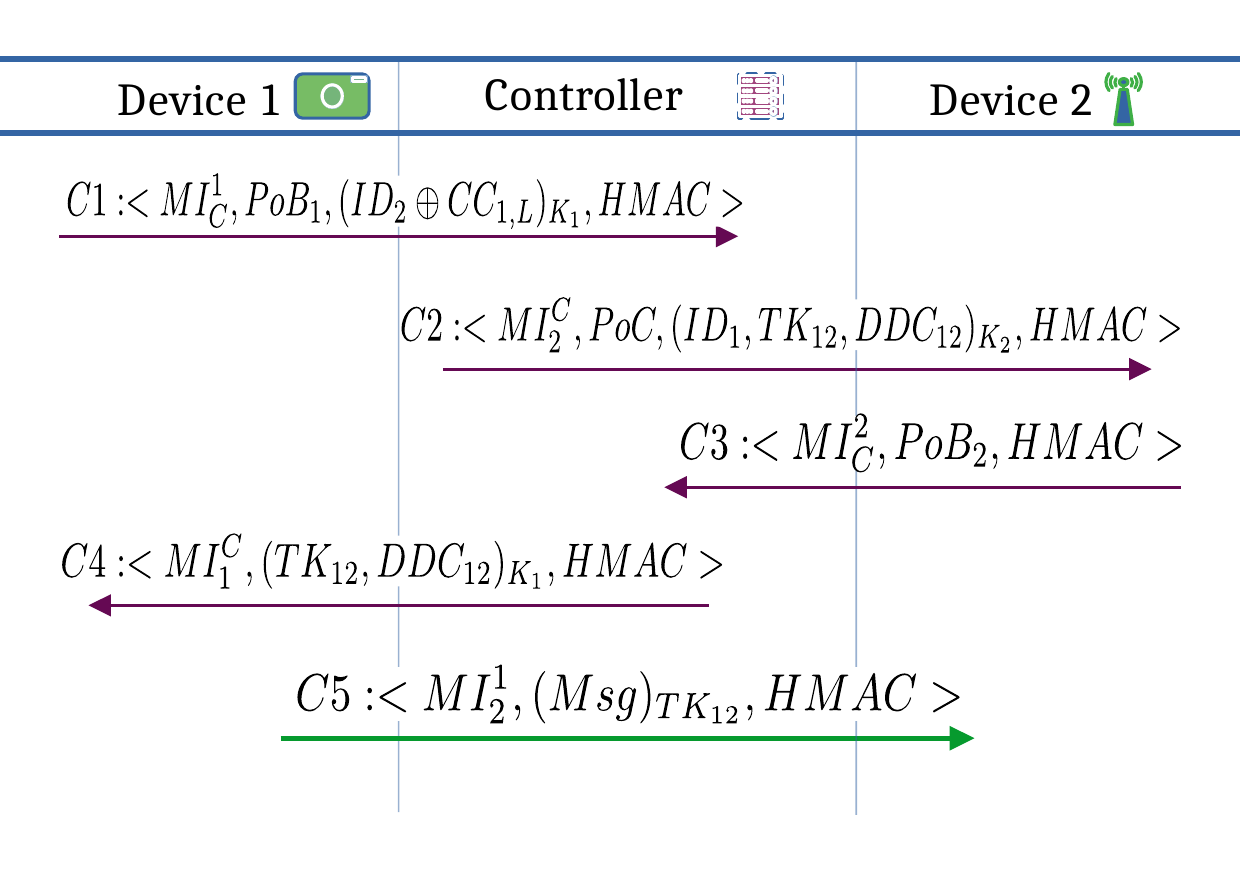}
\caption{Device to Device (D2D) Communication.}
\label{fig3}
\end{figure}

Suppose device $N_1$ wants to communicate with $N_2$, which possesses real identity $ID_{2}$. Both $N_1$ and $N_2$ are authenticated with the controller using the procedure described in Section~\ref{SSC:authentication:phase}. The  D2D communication phase consists of five steps, C1 to C5, which are shown in Fig.~\ref{fig3} and described below.

\textbf{Step C1:} $N_1$ sends the following message to the  controller:
\begin{center}
$<MI^{1}_{C}, PoB_{1}, (ID_{2} \oplus CC_{1,L})_{K_{1}} , HMAC>$, 
\end{center}
where  $MI^{1}_{C}$ is the masked identity of the controller, computed as $h(CC_{1},ID_C)$. $PoB_1 = h(CC_{1}, OTP_{1})$ is the proof of fresh belongingness to the network. $ID_{2}$ is the real identity of $N_2$. $CC_{1,L}$ are the last $L$ bits of $h(CC_1)$, which is the hash of the counter $CC_1$, where $L$ is the length of $ID_2$ in bits. $(ID_2 \oplus CC_{1,L})$ is encrypted using the symmetric key $K_{1}$. XORing $ID_2$ with $CC_{1,L}$ prevents the encrypted output $(ID_{2} \oplus CC_{1,L})_{K_{1}}$ from being the same in the messages sent in step C1 of two different D2D communication exchanges between $N_1$ and $N_2$, with the key $K_1$ being the same in both exchanges. This prevents an eavesdropper from finding out that two different D2D communication exchanges have taken place between the same pair of devices.

HMAC is calculated as $h(CC_{1}, [MI^{1}_{C} \parallel PoB_{1} \parallel (ID_{2} \oplus CC_{1,L})_{K_{1}}])$. After sending this message, $N_1$ increments $CC_{1}$ by 1 and updates $MI^{C}_{1} = h(CC_{1},ID_{1})$ using the fresh value of $CC_1$. $N_1$ will use the updated $MI^{C}_{1}$ to receive the next message from $C$.


\textbf{Step C2:} When the controller receives the message described in step C1, it first calculates $HMAC^* = h(CC_{1},[MI^{1}_{C} \parallel PoB_{1} \parallel (ID_{2} \oplus CC_{1,L})_{K_{1}} ])$  using $CC_{1}$ from its database. If $HMAC^*$ does not match with $HMAC$, then the controller discards the message. 

Next, the controller calculates $PoB_1^* = h(CC_{1}, OTP_{1})$, where $CC_{1}$ and $OTP_{1}$ are taken from its database. If $PoB_{1}$ and $PoB_{1}^*$ match, then this proves the fresh belongingness of $N_1$ to the network and confirms that
successful mutual authentication happened in steps A1 and A2. Next, the controller uses the key $K_{1}$ from its database to extract $(ID_{2} \oplus CC_{1,L})$ and XORs it with $CC_{1,L}$ to obtain $ID_2$.

The controller checks whether $ID_{2}$ is the identity of a legitimate and authenticated device present in its database. Also, dynamic identity removal is facilitated at this step. If $N_2$ was earlier removed from the network, then all message requests coming from it and directed to it will be dropped by the controller at this stage. In this case, the controller informs $N_1$ that $N_2$ is not available for the communication~\footnote{Recall that $C$ also
securely sends an update message to all the connected devices, informing them whenever a new device joins or a device leaves the network.}. 

If $N_2$ is legitimate with key $K_{2}$ and the access control list allows $N_1 \leftrightarrow N_2$ communication, then the controller generates $TK_{12}$, which is the temporary key for inter-device communication between $N_1$ and $N_2$. The controller also generates $DDC_{12}$, which is the device-to-device counter for the masked identity generation in the inter-device communication. The controller sends the following message to $N_2$:
\begin{center}
$<MI^{C}_{2}, PoC, (ID_{1}, TK_{12}, DDC_{12})_{K_{2}},HMAC>$,
\end{center}
where $MI^{C}_{2}$ is the masked identity of device $N_2$ calculated as $h(CC_{2},ID_{2})$. $ID_{1}$ is sent to $N_2$ to make it aware of the requestee, $N_1$, for this inter-device communication. $PoC$, the proof of controller, is provided by the controller to $N_2$ to prove its legitimacy and is calculated as $h(CC_{2}, K_{2}, p_2)$. $ID_1$, $TK_{12}$ and $DDC_{12}$ are encrypted using the symmetric key $K_{2}$ and included in the above message. $HMAC$ here is calculated as $h(CC_{2},[MI^{C}_{2} \parallel PoC \parallel (ID_{1} \parallel TK_{12} \parallel DDC_{12})_{K_{2}}])$ .
 
After sending this message, the controller increments $CC_{2}$ by $1$, calculates $MI^{2}_{C} = h(CC_{2},ID_{2})$ using the incremented value of $CC_2$ and updates $MI^{2}_{C}$ in its database. The controller will use $MI^{2}_{C}$ to receive the next message from $N_2$.


\textbf{Step C3:} 
After receiving the above message sent in step C2, to complete the mutual authentication, $N_2$ sends its proof of belongingness to the controller in the following message:
\begin{center}
$<MI^{2}_{C}, PoB_{2}, HMAC>$,
\end{center}
where $MI^{2}_{C}$ is the masked identity of the controller computed as $h(CC_{2},ID_C)$. $PoB_{2}$ = $h(CC_{2}, OTP_{2})$ is the proof of fresh belongingness to the network from $N_2$. $HMAC$ is calculated as $h(CC_{2},M)$, where $M = [MI^{2}_{C} \parallel  PoB_{2}]$. After sending this message, $N_2$ increments $CC_{2}$ by $1$, calculates $MI^{C}_{2} = h(CC_{2},ID_{2})$ and $MI^{2}_{C} = h(CC_{2},ID_{C})$. $N_2$ will use these identities to communicate with the controller in the next message exchange.


\textbf{Step C4:} At this point, $N_1$ has been authenticated in step C1 by the controller, the controller has been authenticated in step C2 by $N_2$, and $N_2$ has been authenticated in step C3 by the controller.  Now, the controller sends the following message to $N_1$:
\begin{center}
$<MI^{C}_{1}, (TK_{12}, DDC_{12})_{K_1}, HMAC>$, 
\end{center}
where $MI^{C}_{1}$ is the masked identity of $N_1$, calculated as $h(CC_{1},ID_{1})$. The payload is encrypted using $K_{1}$. $HMAC$ is calculated as $h(CC_{1},[MI^{C}_{1} \parallel  (TK_{12} \parallel DDC_{12})_{K_{1}}])$. After sending this message, the controller increments $CC_{1}$ by $1$, calculates $MI^{1}_{C} = h(CC_{1},ID_C)$ and $MI^{C}_{1} = h(CC_{1},ID_{1})$, and updates these in its database. The controller will use these identities the next time it communicates with $N_1$.


\textbf{Step C5:} Upon receiving the message described in step C4, $N_1$ verifies the $HMAC$. It then extracts and stores $TK_{12}$ and $DDC_{12}$ in its database. It also increments the counter $CC_{1}$ by $1$ and updates the values $MI^{1}_{C}$ and $MI^{C}_{1}$ in its database. It sends the following message to $N_2$:
\begin{center}
$<MI^{1}_{2}, (Msg)_{TK_{12}}, HMAC>$,
\end{center}
where $MI^{1}_{2}$ is the masked identity of $N_2$ calculated as $h(DDC_{12},ID_{2})$. $Msg$ is the data message, which $N_1$ intended to send to $N_2$, and is encrypted using $TK_{12}$. $HMAC$ is calculated as $h(DDC_{12},[MI^{1}_{2} \parallel  (Msg)_{TK_{12}}])$.

After this, $N_1$ and $N_2$ can communicate further with each other as long as they want by just incrementing the counter $DDC_{12}$ after each message transmission, deriving new masked identities for each other, and encrypting the payloads using $TK_{12}$. Also, both of them can simultaneously communicate with any other device or the controller without letting them or an eavesdropper know about the ongoing $N_1 \leftrightarrow N_2$ exchanges.

\begin{remark}
During a D2D message exchange, the sender device waits for a fixed time duration to get a response from the receiver. If the sender does not get a response in this time, then it aborts the communication. However, note that since the sender sent a message, which the receiver may not have received, there may be a mismatch in the counter value ($DDC_{12}$) at the sender and receiver. So in the future, if the sender wants to communicate again with the receiver, then it starts again from step C1, so that the counter values $DDC_{12}$ at the sender and receiver are made equal.  
\end{remark}

\begin{remark}
Note that the network model shown in Fig.~\ref{fig1} is a star topology; however, the proposed scheme can be readily adapted to a multihop network, e.g., a ZigBee wireless mesh network, as follows. Suppose a controller and a set of end devices are connected together via a multihop network. When a node (controller or end device) wants to send a message to another node, it broadcasts the message. When a copy of the message reaches the intended recipient, it accepts the message since the masked identity in the receiver field of the message matches an identity in its database. Note that messages are not unicast from the sender node to the receiver node since if that were done, the identity of the receiver node would need to be revealed to one or more of the nodes on the path between the sender and the receiver (since they would need to know which node to forward the message to), thereby violating the anonymity property. It can be easily checked that with the above modification, our proposed scheme achieves all the desired security goals, which are listed in Section~\ref{SSC:security:design:goals}, in a multihop network. 
\end{remark}

\begin{remark}
In the above description of the proposed scheme, we have assumed that there is a secret pre-shared bit string, $p_i$, which is only known to the controller and device $N_i$. An advantage of using a secret pre-shared bit string is that the scheme can be used even when no public-key infrastructure~\cite{granjal2015security} is available. However, in case public-key infrastructure is available, our proposed scheme can be modified to eliminate the need for a secret pre-shared bit string as follows. During registration, device $N_i$ randomly generates strings $p_i$, $K_i$, and $CC_i$, composes the message $(ID_i, p_i, K_i, CC_i)$, attaches a digital signature (which can, e.g., be generated using $N_i$'s RSA private key~\cite{granjal2015security}) to it for message integrity, and sends it to the controller after  encrypting it with the latter's public key (e.g., RSA public key~\cite{granjal2015security}). Alternatively, $N_i$ and the controller execute the Diffie-Hellman key exchange algorithm~\cite{granjal2015security} to agree upon the secret values $(p_i, K_i, CC_i)$. Our proposed scheme achieves all the desired security goals, which are listed in Section~\ref{SSC:security:design:goals}, without the need for a secret pre-shared bit string, if either of the above two methods is used during registration.        
\end{remark}


\section{Security Analysis}
\label{SA}
\subsection{Achieved Security Goals}
\begin{enumerate}[\setlength{\IEEElabelindent}{\dimexpr-\labelwidth-\labelsep}
    \setlength{\itemindent}{\dimexpr\labelwidth+\labelsep}
    \setlength{\listparindent}{\parindent}
  ]
\item{The proposed scheme achieves confidentiality by using symmetric key encryption for all payloads with a dynamic key that gets updated after every session.}
\item{The proposed scheme achieves message integrity by the attachment of an 
$HMAC$, calculated using a counter, to every message transmitted under the scheme, which can be verified by the receiver.}
\item{The proposed scheme achieves anonymity by using dynamic identities for every message exchange.}
\item{The proposed scheme achieves unlinkability since, as explained in the description of the scheme in Section~\ref{PS}, if a node $x$ simultaneously communicates with two different nodes $y$ and $z$, then different dynamic identities are used in the messages exchanged with $y$ and $z$. In particular, as explained in step C5 in Section~\ref{SSC:D2D:communication:phase}, two communicating devices $N_1$ and $N_2$ increment the counter $DDC_{12}$ and derive new masked identities for each other after each message transmission; also, note that if $N_1$ is simultaneously communicating with another device, say $N_3$, then different masked identities are used for the messages exchanged with  $N_2$ and $N_3$. Similarly, the masked identities used in steps A1 and A2 of the authentication phase described in Section~\ref{SSC:authentication:phase} are different for different devices.}
\item{The proposed scheme achieves dynamic entity removal (respectively, addition) with forward (respectively, backward) secrecy by using access control lists, temporary keys for payload encryption, dynamic identities, and updates sent by the controller whenever a new device joins or a device leaves the network.}
\item In the proposed scheme, if an adversary tries to launch a replay attack, he/ she would fail because the recipient would have received that message earlier, increased the counter value and changed the dynamic identity used for reception. 
\item If the adversary can intercept all the traffic exchanged in the network, he/ she would still not be able to find out the identities of the sender and receiver due to the use of dynamic identities  and the fact that the payload is encrypted. Also, the message integrity of all messages is protected using an $HMAC$. Hence, the proposed scheme defends against the Man-in-The-Middle attack. 
\item The dynamic identities are derived by hashing counter values with  real IDs. The symmetric keys are also updated after every session. Since an adversary cannot determine the real identities of the sender and receiver of any message, the proposed scheme defends against offline password, identity, session key and counter  guessing attacks.  
\item{The use of dynamic identities prevents an attacker from impersonating any device since he/ she does not know the counter value used for deriving the required dynamic identity.}
\item{Our scheme does not use any time synchronization and hence is not vulnerable to desynchronization attacks.}
\item{The Sybil attack occurs when a legitimate node uses multiple identities at the same time~\cite{grammatikis2019securing}. In our proposed protocol, all messages are sent using dynamic identities, which are computed using secret counter values; so an attacker is unable to send a message with the sender identity being that of another node.}
\end{enumerate}


\subsection{Comparison with Prior Work}
\label{SSC:comparison:prior:work}
The proposed scheme improves upon schemes proposed in prior work in the following ways:
\begin{enumerate}[\setlength{\IEEElabelindent}{\dimexpr-\labelwidth-\labelsep}
    \setlength{\itemindent}{\dimexpr\labelwidth+\labelsep}
    \setlength{\listparindent}{\parindent}
  ]
\item{In the scheme proposed in~\cite{alshahrani2019anonymous}, public key decryption is required before the verification of $HMAC$ in the first step of the authentication phase. This enables an attacker to launch a DoS attack on  the controller by sending garbage requests. ECC is used in the scheme proposed in~\cite{lohachab2019ecc}, which defends against garbage requests, but takes much more time than symmetric key encryption. Our proposed scheme defends against garbage requests sent to the controller at all stages of the communication. In particular, at step A1,  first $HMAC$ is verified and then the payload is decrypted in our scheme. Also, only symmetric key encryption is used in our proposed scheme, which makes it efficient.}

\item{In our proposed scheme, dynamic identities are used for the controller as well as for all devices, and hence the identities of the controller as well as all devices are anonymous. In contrast, in the scheme proposed in~\cite{alshahrani2019anonymous}, dynamic identities are used only for the devices and not for the controller; so no anonymity is provided for the controller. Also, in the scheme proposed in~\cite{lohachab2019ecc}, fixed identities are used for the controller as well as all devices and hence no anonymity is achieved for the controller or devices.}

\item{Our proposed scheme achieves unlinkability as explained in step C5 in Section~\ref{SSC:D2D:communication:phase}.  In the scheme proposed in~\cite{lohachab2019ecc}, masked identities are not used and hence the unlinkability property is not achieved. In the scheme proposed in~\cite{alshahrani2019anonymous}, masked identities are used, but the same masked identity is used for multiple messages exchanged among a pair of entities and also for messages exchanged by an entity with two different entities; due to this, the unlinkability property is not achieved.}

\item{If the real identity database is exposed in the scheme proposed in~\cite{alshahrani2019anonymous}, an attacker can register with the controller by randomly choosing any key and counter value,  while our proposed scheme defends against this attack by using the shared secret value $p_i$.}

\item{In our proposed scheme, the key $K_{i}$ and the counter $CC_{i}$ are never transmitted between any two devices and they are derived using the initially shared parameter $p_i$. This results in lower overhead in our proposed scheme than in those proposed in~\cite{alshahrani2019anonymous} and~\cite{lohachab2019ecc}, in which the key and counter values are sent over the air.}

\item{In the scheme proposed in~\cite{alshahrani2019anonymous}, an accelerometer reading is used for key generation, which remains static for devices with fixed positions.  In the scheme proposed in~\cite{lohachab2019ecc}, timestamps are used for checking the freshness of messages, which requires overhead for maintaining time synchronization. In our proposed scheme, pseudo-random number generation is used instead of an accelerometer reading for the derivation of the key and counter at a device and the controller; also, no time synchronization is required in our proposed scheme. 
}

\item{Our proposed scheme allows two devices, $N_1$ and $N_2$, to exchange multiple messages with each other after completing steps C1 to C5, without having to perform authentication or key agreement again. 
In the scheme proposed in~\cite{alshahrani2019anonymous}, multiple message exchanges after a given authentication are not allowed, due to the use of a nonce, which is used for a single message only.}

\item{In our proposed scheme, the controller can initiate a communication session with any device that has been inactive after a mutual authentication phase, since the session key and counter are not discarded until the next mutual authentication phase. The above is not possible under the scheme proposed in~\cite{alshahrani2019anonymous}.} 
\end{enumerate}


\section{Numerical Results}
\label{CA}
In this section, we compare the performance of our proposed scheme with those proposed in~\cite{alshahrani2019anonymous} and~\cite{lohachab2019ecc} via numerical computations. The parameter values used to obtain our numerical results are as follows. In our proposed scheme, all the real identities ($ID_i$ and $ID_C$) are assumed to be $16$-bit long~\cite{callaway2002home}. All dynamic identities ($MI$) and keyed hash values ($HMAC$, $PoB$, $PoC$) are $256$-bit long SHA-$256$ hashes~\cite{frankel2003aes}. The counters $CC$, $DDC$ and the random nonce $ns$ used in the scheme proposed in~\cite{alshahrani2019anonymous} are $256$-bit long randomly generated numbers. $OTP$, $p_i$ and $r_i$  are each $16$-bit long, and the symmetric keys $K_{i}$ and $TK_{ij}$ are based on Advanced Encryption Standard (AES) encryption and are $256$-bit long, with the block size of AES being $128$ bits~\cite{frankel2003aes}. $AUTH\_REQ$ is a $2$ bit long string. The length of the encrypted output using AES with a $256$-bit key is calculated as $\mbox{Cipher Length}$ = $(\lceil \mbox{Clear Length}/128 \rceil) * 128$ bits, where $\lceil$  $\rceil$ denotes the ceil function~\cite{frankel2003aes}.

We assume that keyed hashing ($HMAC$) using SHA-$256$ takes $T_H = 460$ ns, encryption or decryption using AES-$256$ of block size $128$ bits takes $T_{Enc} = T_{Dec} = 800$  ns, and asymmetric decryption takes $T_{PubDec} = 0.114$ ms~\cite{kabra2020efficient}.


\begin{figure}[!h] 
\centering
\includegraphics[scale=0.6]{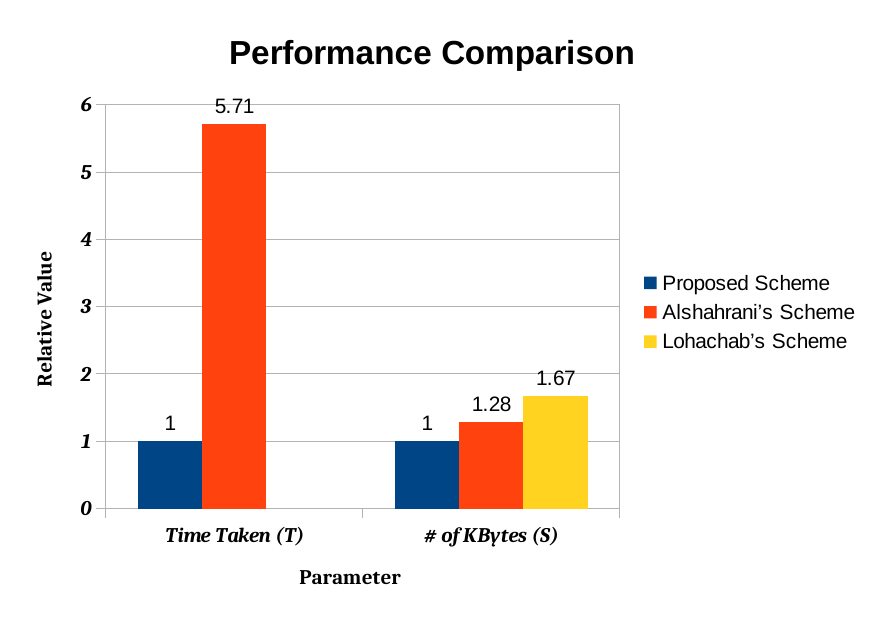}
\caption{The figure compares the performance of our scheme with those of the schemes proposed in~\cite{alshahrani2019anonymous} and~\cite{lohachab2019ecc} (Alshahrani's and Lohachab's schemes, respectively). Note that $T = 24.28$ $\mu$s and $S = 0.76$ Kbytes. The time taken by the scheme proposed in~\cite{lohachab2019ecc} is $3212 \times T$, which is not plotted in the figure since it is much larger than the other values plotted in the figure.}
\label{fig4}
\end{figure}


For the above parameter values, the total time taken for all seven steps of the proposed scheme is $18 \times T_H + 10 \times (T_{Enc} + T_{Dec}) = 24.28$ $\mu$s = $T$. In the scheme proposed in~\cite{alshahrani2019anonymous}, the first step of authentication involves public key cryptography and needs $0.114$ ms and the total time taken for the same payload is $12 \times T_H + 12 \times (T_{Enc} + T_{Dec}) + T_{PubDec} = 138.7$ $\mu$s $= 5.71 \times T$. In~\cite{lohachab2019ecc}, it is stated that the proposed scheme takes $78$ ms for static verification, which is $3212 \times T$. The above comparison is depicted in Fig.~\ref{fig4}.

Our proposed scheme requires $6272$ bits or $0.765$ KBytes $= S$ in total to be transmitted  for a $128$-bit payload delivery between two devices, \emph{with the option of communicating further from either side}, while the scheme proposed in~\cite{alshahrani2019anonymous} takes $8032$ bits or $0.98$ KBytes = $1.28 \times S$ in total for a $128$-bit payload delivery and \emph{no option of further communication}.  The scheme proposed in~\cite{lohachab2019ecc} uses ECC; also, a $256$-bit public and private key pair is used for each communication and it takes a minimum of $1.28$ KBytes = $1.67 \times S$ in total~\footref{note1}. This comparison is shown in Fig.~\ref{fig4}.

Under the proposed scheme, a device needs to store $ID_C$, $CC_i$, $K_i$, $p_i$, $r_i$ and $OTP_i$, i.e., a total of $576$ bits for the controller and $ID_j$, $DDC_{ij}$ and $TK_{ij}$, i.e., a total of $528$ bits for each other device with an active session. The controller needs to store $ID_i$, $CC_i$, $K_i$, $p_i$, $r_i$ and $OTP_i$, i.e., a total of $576$ bits, for each device. The corresponding values for the scheme proposed in~\cite{alshahrani2019anonymous} are $816$, $528$ and $800$, respectively. Also, the correspoding values for the scheme proposed in~\cite{lohachab2019ecc} are $1792$, $1280$ and $1536$, respectively~\footnote{\label{note1}In~\cite{lohachab2019ecc}, no numerical values have been provided for the communication cost or storage at the nodes. The above computations have been performed using the values listed in the first paragraph of Section~\ref{CA}.}.


\section{Conclusions}
\label{CONC}
We presented an authentication and key agreement protocol for a home automation network based on the ZigBee standard. The scheme achieves confidentiality, message integrity, anonymity, unlinkability, forward and backward secrecy, and availability.
Our scheme uses only simple hash and XOR computations and
symmetric key encryption, and hence is resource-efficient. Also, our scheme does not use time synchronization, which requires overhead to achieve. We
showed using a detailed security analysis and numerical results
that our proposed scheme provides better security and anonymity,
and is more efficient in terms of computation time, communication cost, and storage cost than schemes proposed in prior works.


\bibliographystyle{IEEEtran}
\bibliography{refer}

\end{document}